\documentclass[citeautoscript,floatfix,aps,prb,twocolumn,
superscriptaddress]{revtex4-2}

\usepackage{amsmath}
\usepackage{graphicx}
\usepackage{tabularx}
\usepackage{float}
\usepackage{epstopdf}
\usepackage{natbib}
\usepackage{array}
\usepackage{braket}
\usepackage[usenames,dvipsnames]{color}
\usepackage{dcolumn}
\usepackage{bm}
\usepackage{soul} 
\usepackage{changes}
\usepackage{hyperref}
\hypersetup{colorlinks=true, linkcolor=blue, filecolor=blue, urlcolor=blue}
\newcommand{\nv}{NV$^{-}$}
\newcommand{\ci}[1]{C$_{i}^{#1}$}
\newcommand{\vc}[1]{V$_{\text{C}}^{#1}$}

\begin{document}
\title{Density-functional theory study of the interaction between \nv{} centers and native defects in diamond}
\author{Gabriel I. L\'{o}pez-Morales}
\affiliation{Department of Physics and Astronomy, Stony Brook University, Stony Brook, New York, 11794-3800, USA}
\author{Joanna M. Zajac}
\affiliation{Instrumentation Division, Brookhaven National Laboratory, Bldg. 535 P.O. Box 5000, Upton, New York 11973-5000, USA}
\author{Tom Delord}
\affiliation{Department of Physics, City College of the City University of New York, New York, New York 10031, USA}
\author{Carlos A. Meriles}
\affiliation{Department of Physics, City College of the City University of New York, New York, New York 10031, USA}
\affiliation{The Graduate Center of the City University of New York, New York, New York 10016, USA}
\author{Cyrus E. Dreyer}
\affiliation{Department of Physics and Astronomy, Stony Brook University, Stony Brook, New York, 11794-3800, USA}
\affiliation{Center for Computational Quantum Physics, Flatiron Institute, 162 5th Avenue, New York, New York 10010, USA}

\date{\today}

\begin{abstract}
The \nv{} color center in diamond has been demonstrated as a nanoscale sensor for quantum metrology. However, the properties that make it ideal for measuring, e.g., minute electric and magnetic fields also make it sensitive to imperfections in the diamond host. In this work, we quantify the impact of nearby native defects on the many-body states of \nv{}. We combine previous quantum embedding results of strain and electric-field susceptibilities of \nv{} with density-functional theory calculations on native defects. The latter are used to parametrize continuum models in order to extrapolate the effects of native defects up to the micrometer scale. We show that under ideal measuring conditions, the optical properties of \nv{} are measurably affected by the strain caused by single carbon interstitials and vacancies up to 200 nm away; in contrast, the \nv{} is measurably affected by the electric field of such charged (neutral) native defects within a micron (100 nm). Finally, we show how measuring multiple individual \nv{} centers in the vicinity of a native defect can be used to determine the nature of the defect and its charge state.

\end{abstract}
\maketitle
\section{Introduction \label{sec:introduction}}
The negatively-charged nitrogen vacancy (NV$^{-}$) center in diamond is a point-defect color center that has seen an explosion of interest in recent decades due to its promise as an element for quantum and nanoscale technologies. One area where this promise has already been useful is nanoscale metrology, where the \nv{} has been utilized for high-precision measurements of, e.g., strain, electric, and magnetic fields at the nanoscale~\cite{Tamarat2006, Dolde2011, Bassett2011, Olivero2013, Schirhagl2014, Balasubramanian2014, Trusheim2016, Lee2016, Mittiga2018, Kehayias2019, Zhang2021, Bian2021, McCullian2022, Ji2024, Delord2024}. 
The fact that \nv{} and other color centers are so sensitive to their environment also means that they are sensitive to the quality of the host crystal \cite{Doherty2011, Manson2018, Achard2020, Ashfold2020, Barry2020}. Indeed, defects and impurities in the bulk and on the surface significantly degrade the sensing performance of \nv{}~\cite{Doherty2011, Gatto2013, Manson2018, Luhmann2018, Achard2020, Ashfold2020, Yuan2020, Barry2020, Dwyer2022, Neethirajan2023, Kumar2024}. However, this sensitivity could be utilized to, e.g., characterize defects existing in the host from growth, implantation, or radiation damage~\cite{BudnikPRB18, Kirkpatrick2023,Belthangady2013,Sushkov2014,Luhmann2018,Bauch2020,Marcks2024}. For instance, there have even been proposals to detect dark-matter particles via the effects of the damage they cause on \nv{} properties \cite{LukinPRD17,WalsworthAVSQuantumSci22, Dreyer2024}. A key recent experimental development that bolsters this capability is the ability to simultaneously monitor the optical properties of multiple individual \nv{} centers \cite{Chen2019, Ji2024, Delord2024, Guo2024}.

Nearby point defects can affect the properties of \nv{} in several ways. In general, they will cause a strain field that may be anisotropic depending on the structure of the defect and the symmetry of the crystal lattice~\cite{Dudarev2018, Clouet2018,Gengor2024}. Also, the defect will disrupt the periodic lattice potential, and thus result in electric fields; these will be especially important if the defect is charged. The effects of strains and electric fields on \nv{} obey the same symmetry properties, and thus can be difficult to differentiate. Finally, there will be dynamic effects including fluctuations of the spins of paramagnetic defects~\cite{Maze2008}, changes in charge state~\cite{Manson2005}, or even exchange of ionized carriers between defects and \nv{} centers~\cite{Lozovoi2021, Lozovoi2023}. The key challenge to detecting or characterizing point defects via these changes in \nv{} properties is to disentangle all the potential perturbations for (possibly) multiple defects at different locations in crystals that may themselves have built in strains or external electric fields. Thus a comprehensive, quantitative first-principles study of the effects of nearby defects on \nv{} is required, where all of the different contributions can be systematically isolated and explored.

In this work, we use density functional theory (DFT) calculations, combined with previously reported \nv{} susceptibilities from quantum embedding calculations~\cite{Lopez-Morales2024} to quantify how \nv{} centers in diamond are affected by the strain and static electric fields of native defects, and how \nv{} can be used to characterize such defects. We will focus on the simplest cases, that is, isolated carbon interstitials (\ci{}) and vacancies (\vc{}) in the diamond lattice and determine how far away an \nv{} center's many-body energies and thus optical properties are measurably affected. To obtain such effects on the micrometer scale, we parametrize models of the long-range electric/strain field behavior of the defect based on its charge, charge distribution, and elastic dipole. We show that measurable effects on the \nv{} center are caused by the strain of a single defects up to 0.2 $\mu$m away, while electric fields from charged native defects are somewhat longer ranged, causing measurable shifts over a micron away. In addition, we show how determining the effects on multiple individual \nv{} centers can be used to identify the nature and charge state of the nearby defect.

The paper is organized as follows. In Sec.~\ref{sec:methods} we discuss the computational methodology behind the DFT calculations and continuum models. This is followed in Sec.~\ref{sec:results} by a discussion of the results for the strain and electric fields of native defects in diamond, the effects on nearby \nv{} centers, and a discussion of defect characterization vial multi-\nv{} measurements. Finally, we summarize the work and its conclusions in Sec.~\ref{sec:conclusions}.

\section{Computational methodology \label{sec:methods}}
Our \textit{ab-initio} computational approach will proceed in three steps. First, we will determine the long-range strain and electric fields that result from isolated \vc{} and \ci{} in diamond by extrapolating DFT supercell calculations. Then, we will use the strain and electric field susceptibilities calculated in Ref.~\citenum{Lopez-Morales2024} to determine how such defects affect the properties of \nv{}. Finally, we will combine this information with \nv{}-\nv{} interactions (via strain and electric fields) and potential built-in strains to show how multi-\nv{} measurements can probe native defects.

\subsection{Defect-induced strain fields \label{sec:A}}
In order to determine the long-range strain fields of \ci{} and \vc{}, we begin by converging the local relaxations using diamond supercell sizes of 512 (4$\times$4$\times$4), 1000 (5$\times$5$\times$5), and 1728 atoms (6$\times$6$\times$6). Computational parameters of these calculations are given in Sec.~\ref{sec:C}. To extrapolate further, we use an analytical continuum model assuming isotropic elasticity~\cite{Varvenne2017, Dudarev2018, Clouet2018}. In particular, the strain at a point \textbf{r} caused by a defect at the origin is broken down into contributions from unit point-forces given by the isotropic elastic Green's function~\cite{Varvenne2017, Dudarev2018, Clouet2018}
\begin{equation}
G_{ij}(\textbf{r}) = \frac{1}{8\pi \mu r} \left[ \frac{\lambda + 3\mu}{\lambda + 2\mu} \delta_{ij} + \frac{\lambda + \mu}{\lambda + 2\mu} \frac{r_i r_j}{r^2} \right]
\label{eq:1}
\end{equation}

Here, $r = |\textbf{r}|$ and $\delta_{ij}$ is the Kronecker delta. We take the Lamé parameters of diamond, $\lambda$ and $\mu$, as 487.2 and 117.53 GPa, respectively~\cite{McSkimin1972, Fukumoto1990, Guler2015}. The strain can be obtained by summing the second spatial derivatives of the Green's function over the force distribution caused by the defect. Far from the defect, where the induced strain is small, we perform an expansion whose lowest order non-zero term is~\cite{Varvenne2017, Dudarev2018, Clouet2018}
\begin{equation}
\varepsilon_{il}(\textbf{r}) = \frac{1}{2} \sum_{j,k=1}^3 P_{jk} \left( \frac{\partial^2 G_{ij}}{\partial x_k \partial x_l} + \frac{\partial^2 G_{lj}}{\partial x_k \partial x_i} \right)
\label{eq:2}
\end{equation}
The defect-induced elastic dipole $P_{jk}$ can be immediately obtained from a ground-state DFT calculation via
\begin{equation}
P_{jk} = -V \cdot \langle\sigma_{jk}-\sigma_{jk, 0}\rangle,
\label{eq:3}
\end{equation}
where, $\sigma_{jk}$ and $\sigma_{jk, 0}$ represent the average stress in the defective and pristine supercells, respectively, and $V$ the volume of the supercell (kept constant). The elastic dipoles converge faster than the strain at the boundaries of the supercell, thus allowing for extrapolation. While this model specifically describes atomic strain, we find that the $\propto 1/r^{3}$ scaling also accurately extrapolates the decay in, e.g., the local bond length and tetrahedral volume away from the defect. The latter quantities allow for a more direct connection to the \nv{} strain susceptibilities.

We parametrize the effects of strain on the $^{3}E_{x/y}$ excited states of the \nv{} center with the Hamiltonian~\cite{Maze2011, Doherty2011, Lee2016}
\begin{equation}
\label{eq:A1}
\textbf{H}_{\text{strain}} =
\begin{bmatrix}
E + \Delta & \kappa \\
\kappa & E - \Delta
\end{bmatrix}
\end{equation}
in the $\{E_{x}, E_{y}\}$ basis. Here, the rigid shift in the $^{3}E_{x/y}$ manifold, $E$, is given by
\begin{equation}
\label{eq:E}
\begin{split}
E & = E_0 + \chi_{A_1} \varepsilon_{zz} + \chi_{A_1'} (\varepsilon_{xx} + \varepsilon_{yy})
\\
& + \chi_{A_1}^{(2)} \varepsilon_{zz}^2 + \chi_{A_1'}^{(2)} (\varepsilon_{xx} + \varepsilon_{yy})^2;
\end{split}
\end{equation}
where $E_0$ is the zero-strain energy, $\chi$ ($\chi^{(2)}$) are the first (second) order strain susceptibilities, labeled by the irreducible representation of $C_{3v}$ by which they transform ($A_1$ or $E$)~\cite{Lee2016}. The coordinate system for the strains $\varepsilon_{ij}$ is such that $z$ is along the N--V axis, and $x$ points along the $xy$ projection of the $[1\overline{1}1]$ bond. The splitting between $^{3}E_x$ and $^{3}E_y$ is given by
\begin{equation}
\begin{split}
\Delta &= \chi_E (\varepsilon_{yy} - \varepsilon_{xx}) + 2 \chi_{E'} \varepsilon_{xz} 
\\
&+ \chi_E^{(2)} (\varepsilon_{yy} - \varepsilon_{xx})^2 + 2 \chi_{E'}^{(2)} \varepsilon_{xz}^2;
\end{split}
\end{equation}
and the mixing between $^{3}E_x$ and $^{3}E_y$ is
\begin{equation}
\kappa = 2 (\chi_E \varepsilon_{xy} + \chi_{E'} \varepsilon_{yz} + \chi_E^{(2)} \varepsilon_{xy}^2 + \chi_{E'}^{(2)} \varepsilon_{yz}^2).
\label{eq:A4}
\end{equation}

In Sec.~\ref{sec:C2} we will consider multiple \nv{} centers with random orientations. Thus, we will need to rotate the local strain into the frame of the \nv{} in order to use Eqs.~(\ref{eq:A1})--(\ref{eq:A4}).
For a given direction of NV$_z$, we transform the strain via $\bm{\varepsilon}' = \textbf{R} \bm{\varepsilon} \textbf{R}^\top$ where~\cite{Terzakis2018}
\begin{equation}
\mathbf{R} = \mathbf{I} + \bm{v}_\times + \frac{1 - \cos\theta}{\|\bm{v}\|^2} \bm{v}_\times^2
\label{eq:A7}
\end{equation}
and, $\bm{v} = \text{NV}_{z} \times \textbf{z}$, $\cos\theta = \text{NV}_{z} \cdot \textbf{z}$, and $\bm{v}_\times$ is the skew-symmetric matrix of $\bm{v}$.

The strain susceptibility coefficients $\chi_{{}_\Gamma}$ are obtained from the quantum embedding (QE) calculations (see Ref.~\citenum{Lopez-Morales2024}) on strained \nv{}-containing supercells. We consider strain along the \nv{} axis (chosen to be [111]), one of the V--C bonds, i.e., $[1\overline{1}1]$, or perpendicular to NV$_z$, i.e., $[\overline{1}01]$. The shifts and splittings of many-body states for [111] and $[1\overline{1}1]$ strains are reported in Ref.~\citenum{Lopez-Morales2024}, while the $[\overline{1}01]$ case is performed here with identical computational methods as reported in that work. Since we have considered relatively large strains on the \nv supercell, we use the symmetrized Lagrangian form of strain~\cite{Cao2018, Tanner2019}
\begin{equation}
\varepsilon_{ij} = \frac{1}{2} \left( F_{ij}F_{ji} - \delta_{ij}\right),
\label{eq:A6}
\end{equation}
where
\begin{equation}
\mathbf{F} = \mathbf{A} \cdot \mathbf{A}_0^{-1}
\label{eq:A5}
\end{equation}
with $\mathbf{A}$ ($\mathbf{A}_0$) the lattice vectors of the strained (unstrained) \nv{} supercell. The definition of strain in Eq.~(\ref{eq:A6}) is consistent with the strain definitions within our elastic model [i.e., Eqs.~(\ref{eq:2}) and~(\ref{eq:3})]. For the three strains considered,
\begin{subequations}
\begin{equation}
\mathbf{A}_{[111]} = a_0
\begin{bmatrix}
1-\epsilon & -\epsilon & -\epsilon \\
-\epsilon & 1-\epsilon & -\epsilon \\
-\epsilon & -\epsilon & 1-\epsilon
\end{bmatrix}
\end{equation}
\begin{equation}
\mathbf{A}_{[1\overline{1}1]} = a_0
\begin{bmatrix}
1-\epsilon & \epsilon & -\epsilon \\
\epsilon & 1-\epsilon & \epsilon \\
-\epsilon & \epsilon & 1-\epsilon
\end{bmatrix}
\end{equation}
\begin{equation}
\mathbf{A}_{[\overline{1}01]} = a_0
\begin{bmatrix}
1-\epsilon & 0 & \epsilon \\
0 & 1 & 0 \\
\epsilon & 0 & 1-\epsilon
\end{bmatrix},
\end{equation}
\end{subequations}
where, $a_0$ is the equilibrium lattice constant of the supercell and $\epsilon$ is the magnitude of the strain. We found the most robust fitting procedure was to first obtain $\chi_{A_1}$ and $\chi_{A_1}^{(2)}$ from the [111] strain case and $\chi_{E}$ and $\chi_{E}^{(2)}$ from the $[\overline{1}01]$ strain case; then we fixed these values and fitted the rest of the parameters to the $[1\overline{1}1]$ strain calculation. This provided consistent susceptibilities for all strains. 

From this procedure, we obtain coefficients (with quadratic components in parentheses) $\chi_{_{A_1}}$, $\chi_{_{A_1'}}$, $\chi_{_{E}}$, and $\chi_{_{E'}}$ to be 6.48 ($-40.24$), $-7.67$ (17.07), 33.11 ($-82.80$), $-91.88$ (66.12), respectively, all in units of eV. Due to the somewhat large strains considered in Ref.~\citenum{Lopez-Morales2024}, the quadratic components of the $\chi$ coefficients were required to obtain the best fits. However, the strain fields relevant to \nv{}-native-defect interactions are rather weak and well within the linear regime, so the results shown below do not depend strongly on the inclusion of such quadratic components.

The $A_1$ strain susceptibilities were determined experimentally in Ref.~\citenum{Lee2016} to be $\chi_{_{A_1}}=-8.1\pm1.1$ and $\chi_{_{A_1^\prime}}=8.9\pm1.3$ eV. The magnitudes of these values are in quite good agreement with our calculations; we attribute the difference in sign to a different convention in that work, and believe that the trends we find agree. The values for the $E$ coefficients we obtain are one to two orders of magnitude larger than what was reported in Ref.~\citenum{Lee2016}, which we attribute to a different definition of $^{3}E_x$ and $^{3}E_y$ such that there is no avoided crossing at zero strain, see Ref.~\citenum{Lopez-Morales2024} for discussion.

\subsection{Defect-induced electric fields \label{sec:B}}
Local electrostatic fields will be created by dangling bonds or trapped charges on native defects. An important consideration here is that point defects can be thermodynamically stable in different charge states, depending on the Fermi level of the host material. In the dilute limit, however, this Fermi level can be ill-defined due to the tight degree of localization of typical defect states, allowing for local metastable but long-lived non-equilibrium charge dynamics that are particularly prominent in diamond~\cite{Lozovoi2020, Wood2023, Arellano2024, Goldblatt2024}. In such case, prolonged laser illumination of diamond with relatively dilute concentrations of donor/acceptor defects (such as those considered herein) can result in complicated charge dynamics and space-charge effects that can be imaged through nearby \nv{} centers~\cite{Mittiga2018, Delord2024, Ji2024, Goldblatt2024, Li2024}. Thus, we consider all of the stable charge states~\cite{Deák2014, Shim2005} of \vc{} and \ci{} across the diamond bandgap, i.e., $+1$, 0, and $-1$.

The long-ranged electrostatic interactions of point defects has been a significant field of study with the goal of obtaining converged formation and ionization energies \cite{Makov1995,Freysoldt2009, Freysoldt2011, Komsa2013, Kumagi2014, Smart2018, Freysoldt2018, Chagas2021}. In this context, it has been established that the details of the electronic charge density or electrostatic potential of the defect only matters for the short-range interactions, and the long-range electric fields are well described by simple model charge densities~\cite{Freysoldt2009, Freysoldt2011, Freysoldt2018}. Thus our strategy is analogous to the finite-size correction approach employed by Freysoldt \textit{et al.}~\cite{Freysoldt2009, Freysoldt2011, Freysoldt2018}.

We obtain the charge density associated with the defect $\rho_{\text{D}}(\textbf{r})$ by subtracting that of the pristine supercell and performing a macroscopic average of the resulting charge density. For atoms added/removed to create the defect, we add/remove Gaussian charges at their approximate locations, with the width chosen to be the projector-augmented wave (PAW) radius of the atom (see Sec.~\ref{sec:C}). Thus, the ``excess'' charge density $\rho_{\text{D}}(\textbf{r})$ integrates to the charge state of the defect.

\begin{figure*}[h t]
\centering
\includegraphics[width=1.0\textwidth]{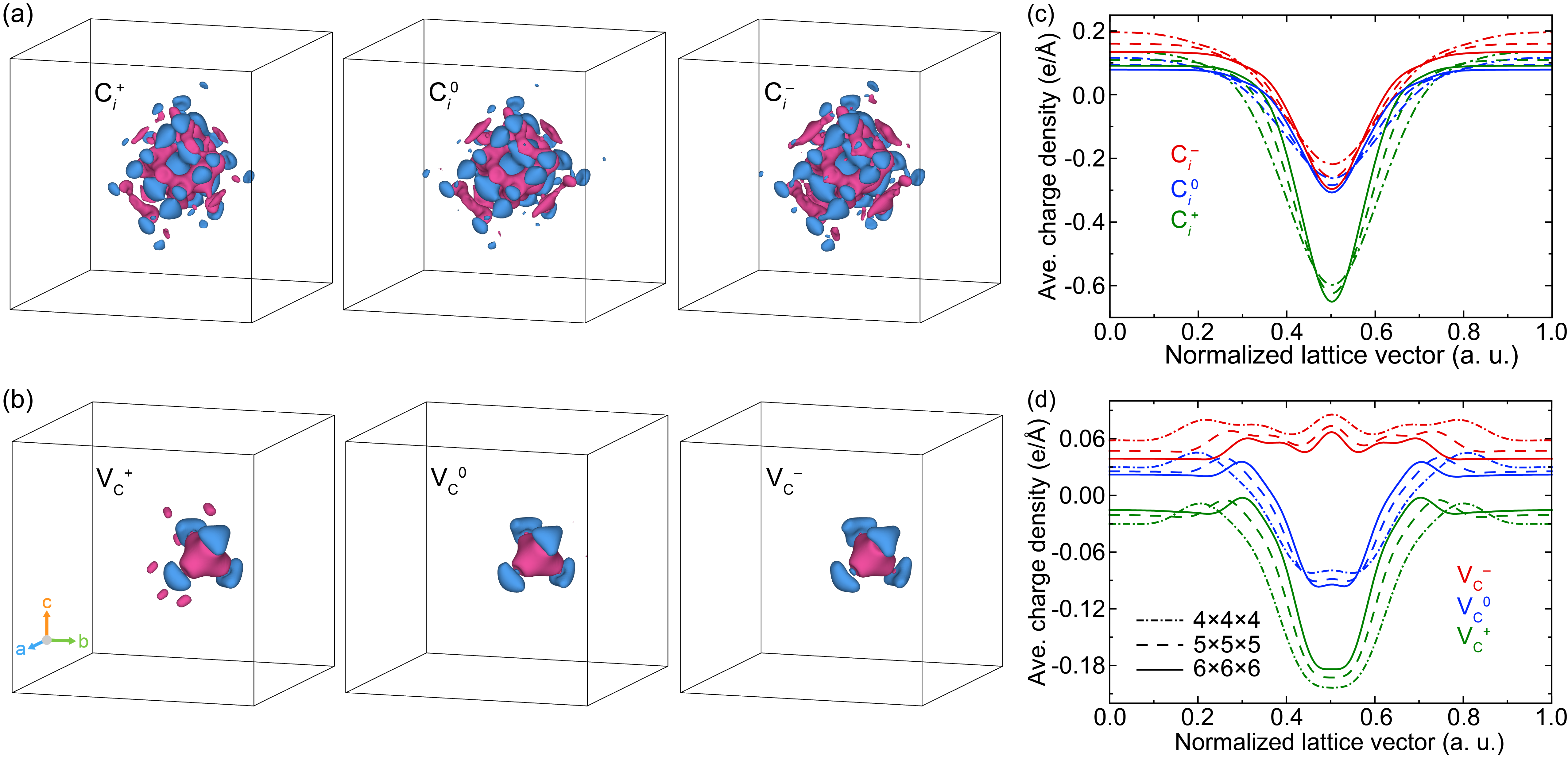}
\caption{Defect charge densities for intrinsic defects in diamond. (a, b) Excess defect charge densities for the \ci{} and \vc{} defects (respectively) in the $4\times4\times4$ supercell for the three most-stable charge states. Pink/blue shades correspond to positive/negative iso-surfaces. All iso-surfaces are set to 0.12 e/\AA. (c, d) Macroscopic average of the excess defect charge densities exemplified in (a, b), respectively, for different supercell sizes.}
\label{fig:1}
\end{figure*}

Using these defect charge densities, we extrapolate the macroscopic electric fields at arbitrary distances away from the defects using a multipole expansion of the Coulomb field. Specifically, 
\begin{equation}
\mathbf{E}(\textbf{r}) = \frac{1}{4\pi\epsilon r^{2}} \cdot [Q_{\text{D}} + \frac{1}{r}(3(\textbf{p}_{\text{D}} \cdot \hat{\textbf{r}}) \hat{\textbf{r}} - \textbf{p}_{\text{D}})]
\label{eq:4}
\end{equation}
where, $Q_{\text{D}} = \int\rho_{\text{D}}(\textbf{r})d\textbf{r}$ and $\textbf{p}_{\text{D}} = \int \textbf{r} \cdot \rho_{\text{D}}(\textbf{r})d\textbf{r}$ are the charge monopole and dipole moments of the defect, respectively. Note that for the charged cases, the dipole moment will depend on the choice of origin; we take the position of maximum defect charge as the origin, which for somewhat symmetric defects allows the dipolar part to capture the deviation from spherical symmetry of the charge density. Here, $\epsilon$ is the dielectric constant of diamond (taken to be $5.7\epsilon_{0}$~\cite{Bhagavantam1948, Ibarra1994}). Overall, this approach allows to approximately capture short-range effects due to the local spatial distribution of the defect charge densities; as we show below, the only interactions over any appreciable distances are the monopole ones.

The effects of the electric field on the \nv{} are determined, again, using the QE calculations of Ref.~\citenum{Lopez-Morales2024}. In that work, the permanent dipole moments of the $^{3}E_{x/y}$ many-body states (referred to the $A_2$ ground state, and thus relevant for the optical transition), and the transition dipole moment between them were determined. Similar to the case of strain, electric fields parallel to the NV$_z$ shift both $^{3}E_x$ and $^{3}E_y$ together, while perpendicular fields will split $^{3}E_x$ and $^{3}E_y$. Thus the electric field Hamiltonian in the basis \footnote{Note that $^{3}E_x$ and $^{3}E_y$ are chosen to be eigenstates in the point group $C_s$, i.e., after the symmetry is broken by field or strain. The $x$ direction is taken to be in the remaining mirror plane, and $y$ perpendicular to it.} where NV$_z$ is along the cartesian $z$ can be written as~\cite{Maze2011}
\begin{equation}
\label{eq:el_field}
\textbf{H}_{\text{electric}} = d_\parallel
\begin{bmatrix}
\mathcal{E}_z & 0 \\
0 & \mathcal{E}_z
\end{bmatrix}+d_\perp
\begin{bmatrix}
\mathcal{E}_x & \mathcal{E}_y \\
\mathcal{E}_y & -\mathcal{E}_x
\end{bmatrix}
\end{equation}
where $\bm{\mathcal{E}}$ is the electric field. The results in Ref.~\citenum{Lopez-Morales2024} were presented for an \nv{} oriented along the [111]; if we rotate it such that NV$_z$ is along the Cartesian $z$ direction, we find dipole moments characterized by $d_\parallel=1.63$ and $d_\perp=2.16$ D~\cite{Delord2024-2}.

Note that the directions of the dipole moments in the direction perpendicular to the NV$_z$ are a matter of basis choice under $C_{3v}$ symmetry. However, as discussed below, most \nv{} centers in real samples are under some amount of intrinsic transverse strain that slightly breaks the $^{3}E_{x/y}$ degeneracy, and this will fix the orientation of the permanent dipole moments~\cite{Lopez-Morales2024}; thus we will use the local strain to set these directions. If we assume that the \nv{} under strain has $C_s$ symmetry, then the strain direction will also determine the polarization direction for, e.g., the transition between $^{3}E_x$ and $^{3}E_y$. In this case, to cause a transition, the microwave (MW) field must have a component perpendicular to the mirror plane containing the NV$_z$ and the strain direction. In terms of optical transitions from $^3A_2$ to individual $^{3}E_x$ and $^{3}E_y$ states, one transition (to the excited state that transforms like $A'$ under $C_s$) will follow the same selection rules as between $^{3}E_x$ and $^{3}E_y$, while the other (to the excited state that transforms like $A''$) will require a polarization component in the mirror plane to cause a transition.

\subsection{Multi-\nv{} models \label{sec:multi_NV}}
Using the previously-discussed continuum models, we can simulate the strain and electric-field effects of a native defect on multiple nearby \nv{} centers. To demonstrate this, we will create a hypothetical random (in position and orientation) distribution of \nv{} centers around a single native defect in diamond. We assume access to the full spectroscopic degrees of freedom of the \nv{} center's excited $^3E$ manifold (e.g., excitation frequency and polarization), which allow to sense not only the transverse and longitudinal local strains and electric fields but also the orientation of the transition dipoles. In practice, such measurements have been achieved by rotating the optical excitation electric fields ~\cite{Lee2016, McCullian2022}, or characterizing the MW excitation electric field via multi-\nv{} measurements and simulating the MW antenna.

The \nv{} concentration will be taken to be 12.5 NV/$\mu$m$^{3}$, which, assuming a 10\% conversion efficiency, corresponds to $\sim 0.7$ parts per billion (ppb) of nitrogen. This falls within typical concentrations for electronic grade diamond crystals, and translates into $\sim10$ NV centers within a $2\times2\times0.2$ $\mu$m$^{2}$ excitation volume. We distribute and orient the \nv{} centers at random within this volume, assuming a single native defect (e.g., \ci{} or \vc{}) to be at its center. Using the effective models of strain and electric fields described in Secs.~\ref{sec:A} and~\ref{sec:B}, we calculate the net $^{3}E_{y/x}$ spectral shifts/splitting for each \nv{} across the random distribution. We also include NV--NV interactions using the same methodology. Lastly, we consider an arbitrary homogeneous strain field across the NV distribution to ``mimic'' the effects of mesoscopic background or bias fields inevitably present in real experiments~\cite{Lee2016, McCullian2022, McCullian2024, Delord2024, Ji2024}.

An additional spectroscopic resource to the shifts/splitting of the $^{3}E_{y/x}$ levels is the dependence of the strength of the transition between split $^{3}E_{y}$ and $^{3}E_{x}$ excited states of the \nv{} on the polarization direction of the MW field. As previously stated, we calculate the alignment of the $\bm{\mu}_{xy}$ based on the net local strain, with the underlying assumption that strain is always present in the sample. With the NV$_{z}$ axes assigned at random, we exploit the selection rules derived in Refs.~\citenum{Doherty2013, Lopez-Morales2024}, i.e., that $^{3}E_{x} \leftrightarrow$ $^{3}E_{y}$ transitions require a component of the excitation field perpendicular to the mirror plane of the \nv{}; this plane is the one which contains the strain/field direction and the NV$_{z}$ axes. Thus, the orientation of the transition dipole moment is $\bm{\mu}_{xy} \propto \bm{\varepsilon}(\textbf{r}) \times \text{NV}_{z}$, with $|\bm{\mu}_{xy}| = 2.2$ D~\cite{Lopez-Morales2024}. The strength of the transition between the states will be proportional to the square of the projection of the field polarization onto $\bm{\mu}_{xy}$. We note that this should be considered an upper bound for the polarization of the transition, which is only achieved if $\bm{\varepsilon}(\textbf{r})$ is in a high enough symmetry direction so that the resulting symmetry of the \nv{} after perturbation is $C_s$ (i.e., one of the mirror planes survives).

\subsection{Computational parameters \label{sec:C}}
All DFT calculations are performed using the VASP code~\cite{Kresse1996}, with PAW pseudopotentials to treat core electrons~\cite{Bloch1994}. The valence electrons are expanded in a plane-wave basis, with exchange-correlation interactions described via the semi-local generalized-gradient approximation (GGA) parametrized by Perdew, Burke and Ernzerhof (PBE)~\cite{Perdew1996}. A kinetic energy cutoff of 500 eV is chosen for the plane-wave expansion in all cases. We sampled the Brillouin zone at the $\Gamma$-point. All calculations make use of tight convergence criteria for forces (0.001 eV/\AA{}) and electronic iterations (10$^{-8}$ eV) to ensure well-converged atomic structures and wave functions. Using these parameters, we find an ideal bond length of 1.547 \AA{} for the pristine diamond cell, which is used to evaluate all lattice distortion due to the intrinsic defects studied herein. For the case of \ci{}, we consider the ``split-interstitial'' configuration, where the additional carbon atom forms a dimer-like bond with the closest carbon site. We find this to be the most stable configuration within our calculations.

\section{Results \label{sec:results}}
\subsection{Effect of defect-induced electric fields on \nv{}} \label{sec:A2}

\begin{figure*}[h t]
\centering
\includegraphics[width=0.9\textwidth]{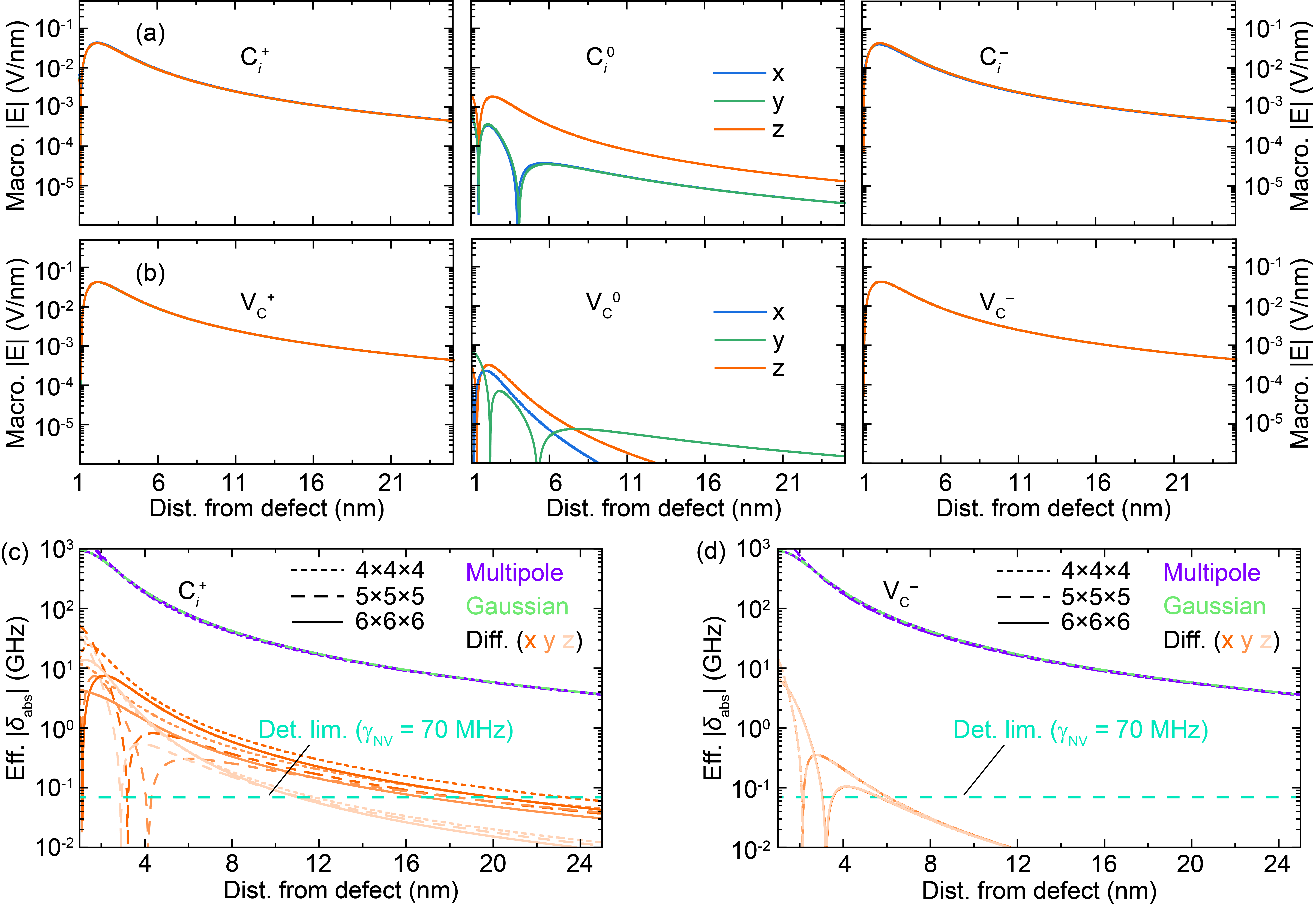}
\caption{Near-field electric-field-induced effects on the \nv{} in diamond as basis for discriminating proximal defect-related charge-traps. (a) Left to right: Macroscopic electric fields generated by \ci{+}, \ci{0}, and \ci{-} within a $\sim$25-nm radius. Lines are color-coded with the three Cartesian directions. (b) Same as in (a), but for the case of \vc{+}, \vc{0}, and \vc{-}. (c, d) Effective energy shifts on the vertical excitation of the \nv{} ($\delta_{\text{abs}}$) from \ci{} and \vc{} (respectively) in different supercells, based on the norm of the multipole electric fields (purple lines) in (a) and (b), compared against a spherically-symmetric Gaussian model (green lines). The differences between these two models along different directions are also included (color-coded tones of orange). All results in (c, d) are based on a permanent dipole of 1.64 D~\cite{Lopez-Morales2024}. Horizontal dashed lines represent the detection limit based on a 70 MHz optical linewidth (inhomogeneous-linewidth dominated).}
\label{fig:2}
\end{figure*}

Here, we consider the energy shifts on the \nv{} arising from the electric fields generated from \ci{} and \vc{}, either charged or neutral, and the distances at which this would be optically detectable. Since we are focusing on long-ranged electrostatic interactions, care must be taken in ensuring convergence of the results within DFT periodic supercells. Hence, we start by comparing the excess charge densities for the defects for the different defect supercells for both \ci{} and \vc{}. As shown in Figs.~\ref{fig:1}{(a) and (b)}, the densities for the defects considered here are relatively well localized to their atomic centers. Overall, those for the \ci{} extend farther than those for the \vc{}, suggesting a more delocalized electronic state in the former case. Just as in the case of the electrostatic DFT potentials, the defect charge at the cell boundaries converges rather slowly with supercell size~\cite{Freysoldt2009, Freysoldt2011, Freysoldt2018}. This is true even for tightly localized defects (as those considered here), and is due to a combination of the periodic boundary conditions and the long-range nature of the electrostatic interactions. However, we see from Figs.~\ref{fig:1}{(c) and (d)} that in both \ci{} and \vc{} defects the relative changes between the two largest supercells remain relatively small ($<12\%$) for all charge states, showing decent convergence of the defect densities in our largest supercells (i.e., $5\times5\times5$ and $6\times6\times6$). From these $6\times6\times6$ defect densities we obtain $\textbf{p}_{\text{D}} = 0.7, 0.2, 0.2$ $e\cdot$\AA{} for \ci{0} (with split-interstitial axis along $x$ axis), and $\textbf{p}_{\text{D}} = 0.06, -0.05, -0.05$ $e\cdot$\AA{} for \vc{0}.

In Fig.~\ref{fig:2} we show the results of the extrapolated electric fields generated by the excess charge densities (Fig.~\ref{fig:1}) in the vicinity ($<30$ nm) of the native defects. Panels (a) and (b) present the different Cartesian components of the field for \ci{} and \vc{}, respectively. Not surprisingly, we see that the field of charged defects is dominated by the spherically symmetric monopole contributions. Of course the electric fields from +/- charged defects will differ in sign, which is neglected here for plotting purposes, and could be used to discriminate between, e.g., \ci{+} and \vc{-} defects. For the neutral cases, significant anisotropy arising from the dipole terms is present, but the magnitudes of the fields are orders of magnitude smaller than the charged cases. Provided the \nv{} is sufficiently close to sense their dipolar (i.e., $\propto 1/r^{3}$, discussed below) electric fields, it should be possible to discriminate between neutral \ci{} and \vc{} defects solely based on these features.

\begin{table*}[ht]
\setlength{\tabcolsep}{1pt}
\renewcommand{\arraystretch}{1.5}
\caption{Elastic dipole elements $P_{ij}$ for all defect centers relevant to this work. The elastic dipole for the \nv{} center was obtained from the $4\times4\times4$ supercell with NV$_z$ along the [111]. All quantities are in $\text{GPa}\cdot\text{nm}^{3}$ units.}
\label{table:1}
\centering
\begin{tabularx}{\textwidth}{c*{7}{>{\centering\arraybackslash}X}}
\hline\hline
Defect & $P_{xx}$ & $P_{yy}$ & $P_{zz}$ & $P_{xy}$ & $P_{xz}$ & $P_{yz}$ \\
\hline\hline
\ci{+} & $-6.066$ & $-0.750$ & $-1.596$ & 0 & $9.86\times10^{-5}$ & $-2.96\times10^{-5}$ \\
\hline
\ci{0} & $-9.072$ & $-2.502$ & $-2.502$ & $3.94\times10^{-5}$ & $-3.94\times10^{-5}$ & $1.97\times10^{-5}$ \\
\hline
\ci{-} & $-12.263$ & $-3.368$ & $-5.649$ & $-4.93\times10^{-5}$ & $-8.87\times10^{-5}$ & $-2.96\times10^{-5}$ \\
\hline
\vc{+} & 1.800 & 2.504 & 1.824 & 0 & 0.593 & $-2.96\times10^{-5}$ \\
\hline
\vc{0} & $-0.696$ & $-0.696$ & 0.827 & $-1.97\times10^{-5}$ & $3.11\times10^{-3}$ & $-2.14\times10^{-2}$ \\
\hline
\vc{-} & $-2.585$ & $-2.585$ & $-2.585$ & 0 & 0 & 0 \\
\hline
\nv & $-2.288$ & $-2.288$ & $-2.288$ & $-0.596$ & $-0.596$ & $-0.596$ \\
\hline\hline
\end{tabularx}
\end{table*}

To further highlight these insights, we use the \emph{average permanent dipole} difference between $^3A_2$ and $^3E$ states of the \nv ($\Delta p = 1.64$ D)~\cite{Lopez-Morales2024} to translate the electric fields into corresponding energy shifts induced on the vertical excitation of the \nv{} ($\delta_{\text{abs}}$). This corresponds to the case where the sensing protocols are employed to measure \emph{only} the longitudinal component of the shifts. Additionally, we present the dipole contribution via the directional difference between $\delta_{\text{abs}}$ obtained through the multipole expansion with that obtained from a Gaussian spherical charge [Figs.~\ref{fig:2}{(c) and (d)}]. We add a detection limit for the \nv{} based on the typical (70-MHz) optical inhomogeneous linewidth as a reference~\cite{Chakravarthi2021, Ji2024, Delord2024}. (Note that this linewidth is mainly caused by spectral diffusion from fluctuations of surrounding defect charge states, and may be reduced with proper protocols \cite{Delord2024, Li2024}, as discussed below.) Exemplifying the case of \ci{+} versus \vc{-}, we see that the directional asymmetries in the generated electric fields [orange-colored lines in Figs.~\ref{fig:2}{(c) and (d)}] extend a few nanometers farther for \ci{}, consistent with the charge delocalization in \ci{} relative to \vc{} described earlier [Figs.~\ref{fig:1}{(a) and (b)}]. Overall, the asymmetries remain quantifiable by an \nv cluster within a radius 5--20 nm away from the native defects; within this radius, the short-range features of the defect charge densities could be used to discriminate between them. These conclusions are consistent across different supercells, highlighting the convergence of our extrapolated results (we focus on the $6\times6\times6$ supercell for the remainder of this section). Though we only illustrate discrimination between \ci{+} and \vc{-} (for brevity), we find these trends to be qualitatively consistent between \ci{-} and \vc{+}, as well as for \ci{0} and \vc{0}. Lastly, we note that sensing of these short-range directionalities from the defect electric fields could be substantially improved by working under resonant excitation at cryogenic temperatures, due to enhanced directional sensing of the \nv{} under such conditions~\cite{Lopez-Morales2024, Delord2024, Delord2024-2, Ji2024}.

\begin{figure}[h b]
\centering
\includegraphics[width=0.75\linewidth]{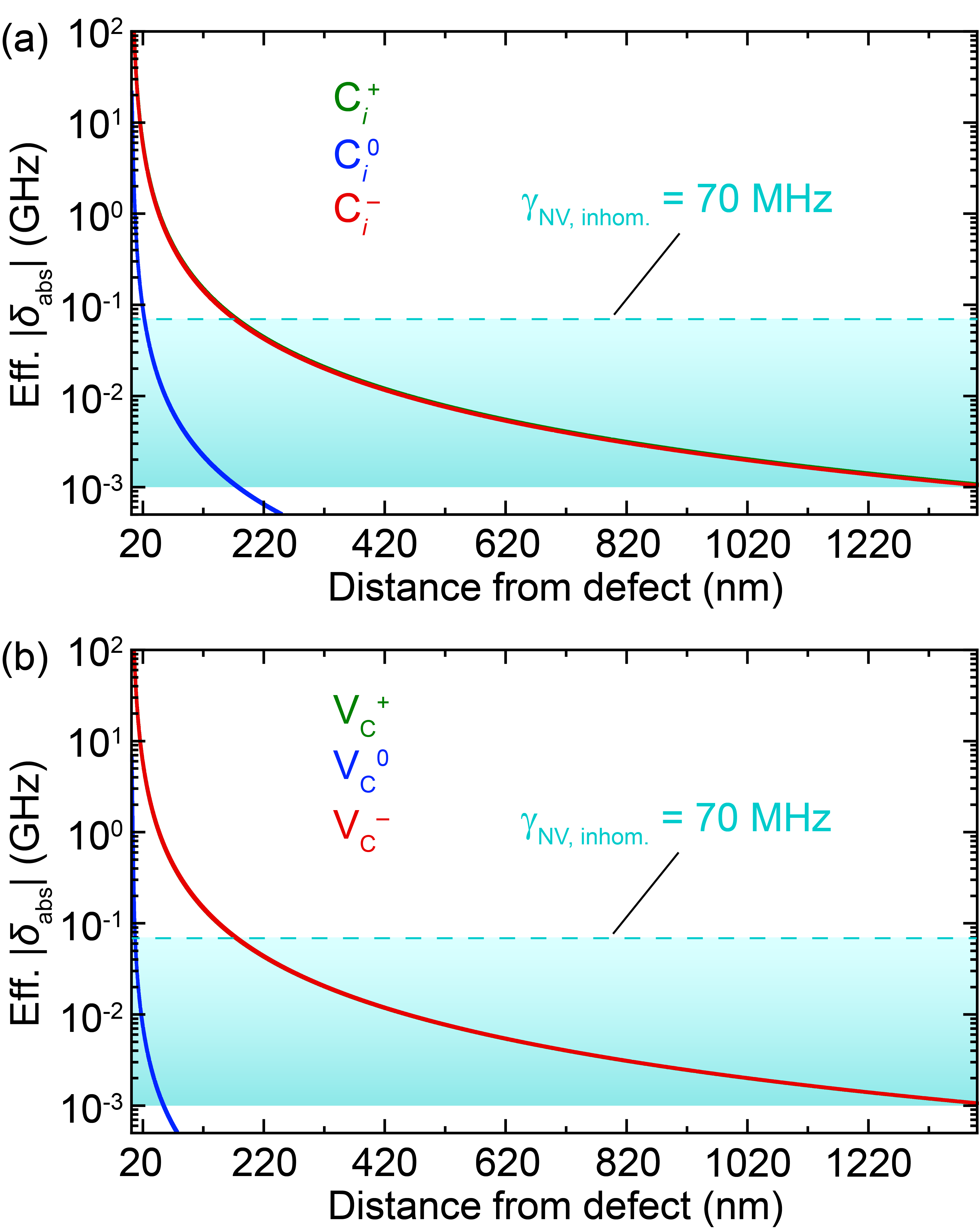}
\caption{\nv{} optical detection limits for single intrinsic defects in diamond based on electric-field-induced energy shifts. (a, b) Electric-field-induced $\delta_{\text{abs}}$ from \ci{} and \vc{} defects, extrapolated to the far field. The results from different supercells are indistinguishable at this scale for all the different charge states, and thus not included. Horizontal dashed lines represent the detection limit based on the typical inhomogeneous (70 MHz) optical linewidth. Shaded regions highlight the resolution range based on experimental detection protocols reported in the literature~\cite{Ji2024}.}
\label{fig:3}
\end{figure}

We present $\delta_{\text{abs}}$ extended to the far field in Fig.~\ref{fig:3}. In this case we are interested in quantifying how far an \nv{} center can reliable sense the generated electric fields. As previously stated, \nv{} centers in high purity diamond have typical inhomogeneous linewidths of around 70 MHz~\cite{Chakravarthi2021, Ji2024, Delord2024}. We note, however, that the actual resolution experimentally achievable depends on the protocol employed, and can greatly exceed the inhomogeneous linewidth. Most notably, electric field fluctuations can be suppressed by neutralization of surrounding charge traps, leading to linewdith below 20 MHz, and time averaging further lowers the detection limits; from Ref.~\citenum{Ji2024} we anticipate frequency sensitivities of 27 and 8 MHz/$\sqrt{\text{Hz}}$ for 70 and 20 MHz linewdiths respectively. Hence, to provide realistic (but conservative) estimates, we highlight detection-limit ranges based on \nv{} linewdiths between 1--70 MHz (shaded regions in Fig.~\ref{fig:3}) to reflect potential experimental enhancement of the detection limits. From these results, we find that the maximum detection radius ranges from 180 nm up to 1.4 $\mu$m for the charged \ci{} and \vc{} defects, and 20--180 (10--50) nm for neutral \ci{} (\vc{}). The fact that the negatively- and positively-charged cases are identical confirm that these two behave the same (as point charges) in the far field, though again, they would have different signs. 

Overall, our extrapolated energy shifts suggest that the \nv{} can reliably sense charged \ci{} and \vc{} defects within a radius of 0.2--1.4 $\mu$m. We point out that resolving the individual permanent dipoles of the $^{3}E_{y/x}$ states at low temperatures could improve this detection limit via narrower linewdiths and slightly larger permanent dipoles~\cite{Lopez-Morales2024, Delord2024, Delord2024-2, Ji2024}. Lastly, the access to additional information under resonant excitation of the $^{3}E_{y/x}$ states, such as local orientation of the fields, could allow to discriminate between the different defect variants even at such large distances.

\begin{figure*}[h t]
\centering
\includegraphics[width=1.0\textwidth]{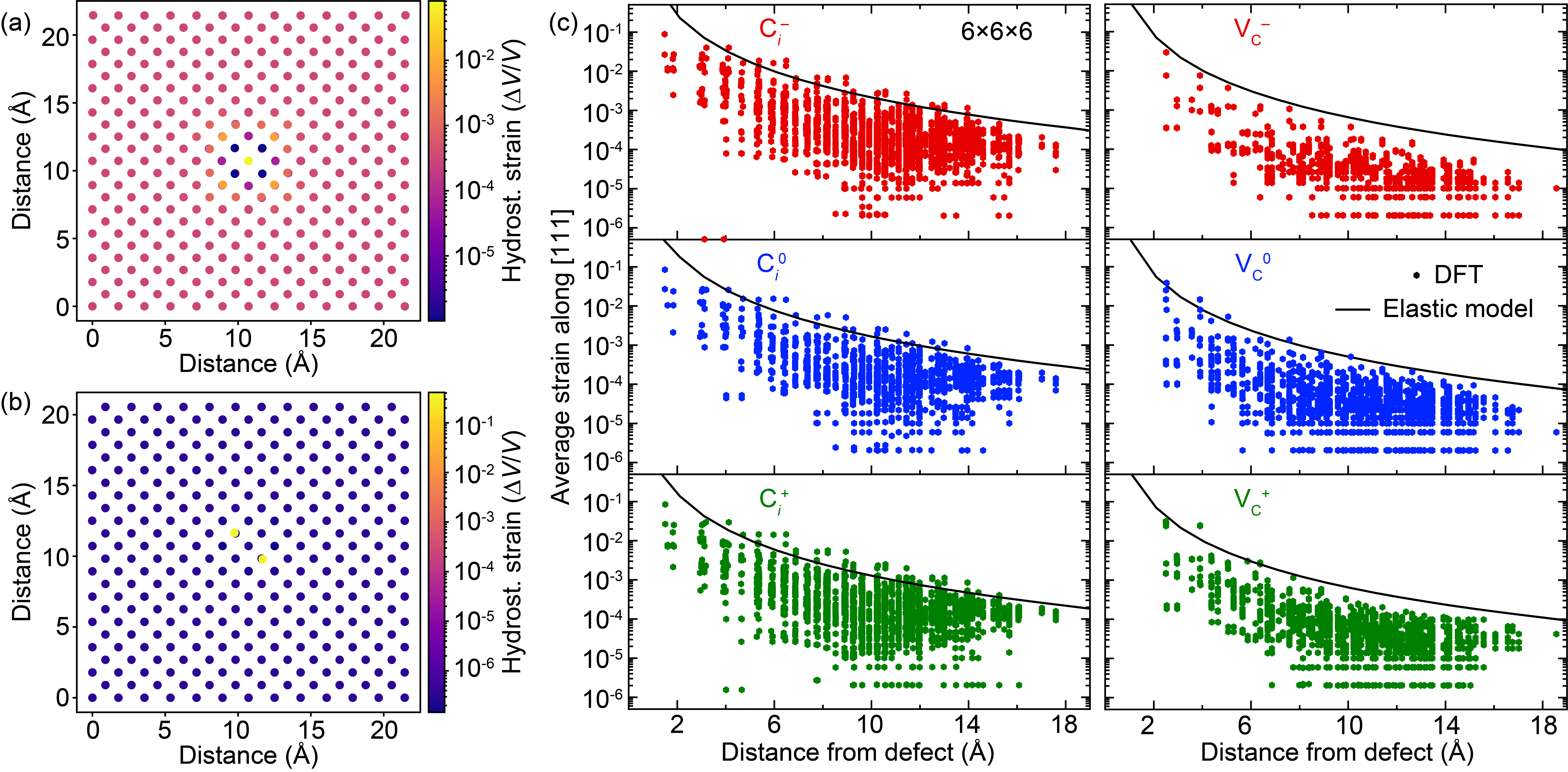}
\caption{Defect-induced strain and displacement fields in diamond from individual \ci{} and \vc{} centers. (a, b) Hydrostatic strains generated by the neutral \ci{} and \vc{} defects (respectively) in a $6\times6\times6$ diamond supercell. (c) Average strain ($\Delta L_{B}/L_{B}$) along the [111] direction for three charge states (top to bottom) of the \ci{} and \vc{} defects (left, right panels, respectively). The solid lines represent the strain calculated via Eq.~\ref{eq:2}.}
\label{fig:4}
\end{figure*}

\subsection{Effect of defect-induced strain fields on \nv{} \label{sec:B2}}
In this section, we turn to the case of strain-induced shifts of the \nv{} optical transitions. The \nv{} strain susceptibilities~\cite{Maze2011, Doherty2011, Lee2016, Lopez-Morales2024} were discussed in Sec~\ref{sec:A}, and the remaining ingredient is the strain profile of the native defects.

A summary of the obtained results is presented in Fig.~\ref{fig:4}. In Figs.~\ref{fig:4}{(a) and (b)}, we plot the local hydrostatic strain-induced by neutral \ci{} and \vc{} defects, respectively, in a slice perpendicular to the Cartesian $z$ direction of the supercell containing the defect. Here, the neutral defect charge states are chosen as examples. Of course, the largest strain is localized at the defect site, with a relatively sharp decay partly driven by the finite size of the defect supercell, i.e., by symmetry constrains at the cell boundaries. The strain appears to propagate slightly longer distances for the interstitial case, resulting in a larger residual strain at the boundaries. This is reasonable, since an interstitial atom should introduce a larger local lattice distortion compared to a missing atom in the lattice. 

To better quantify the atomic distortions induced by the native defects, we gather the local bond-length changes $\Delta L_{B}/L_{B}$, representing defect-induced atomic strains, versus distance from the defect for all lattice points across the $6\times6\times6$ supercells. The results are presented by the points in Fig.~\ref{fig:4}{(c)}, color-coded with their respective defect charge states. The relative changes in local deformation due to the different charge states of a given native defect remain relatively small, while both initial and residual strain at the boundaries are indeed larger for the \ci{} defect. Overall, the strain fields underlying the results summarized in Figs.~\ref{fig:4}{(a--c)} are consistent with recent work employing similar DFT methods~\cite{Kirkpatrick2023}.

In addition to the strain directly obtained from DFT, we include in Fig.~\ref{fig:4}{(c)} the decay predicted by the effective model of strain based on the elastic dipole induced by the point defects in question~\cite{Varvenne2017, Clouet2018, Dudarev2018} (see Sec.~\ref{sec:A}). In Table~\ref{table:1} we give the values of the calculated elastic dipoles for the native defects, as well as the \nv{} itself. The effective decay of strain within this elastic model follows a $\propto1/4\pi\mu\textit{r}^{3}$ dependence, where $\mu$ is the shear modulus of diamond~\cite{Clouet2018}. Comparing these results, we see that the elastic dipole model (solid black lines in Fig.~\ref{fig:4}{(c)}) well describes the general trend of the decay, and provides an upper bound to the DFT strains across the supercell, due to the fact that the latter is limited by the periodic boundary conditions.

This comparison demonstrates the validity of the elastic model for extrapolating $\delta_{\text{abs}}$ at distances realistic for sensing experiments (e.g., $> 100$ nm). Another advantage of this approach is that the convergence of the elastic dipole with supercell size is typically much faster than the residual strain at the cell boundaries. In fact, we find that the elastic dipoles $P_{jk}$ are already converged at the $4\times4\times4$ cell size for both \ci{} and \vc{}, while the strain at the boundary follows a linear decay with supercell size for the \ci{}. In all the results that follow, we use the elastic dipoles $P_{jk}$ obtained for all relevant defects through the methodology in Sec.~\ref{sec:A} (Table~\ref{table:1}).

\begin{figure}
\centering
\includegraphics[width=1.0\linewidth]{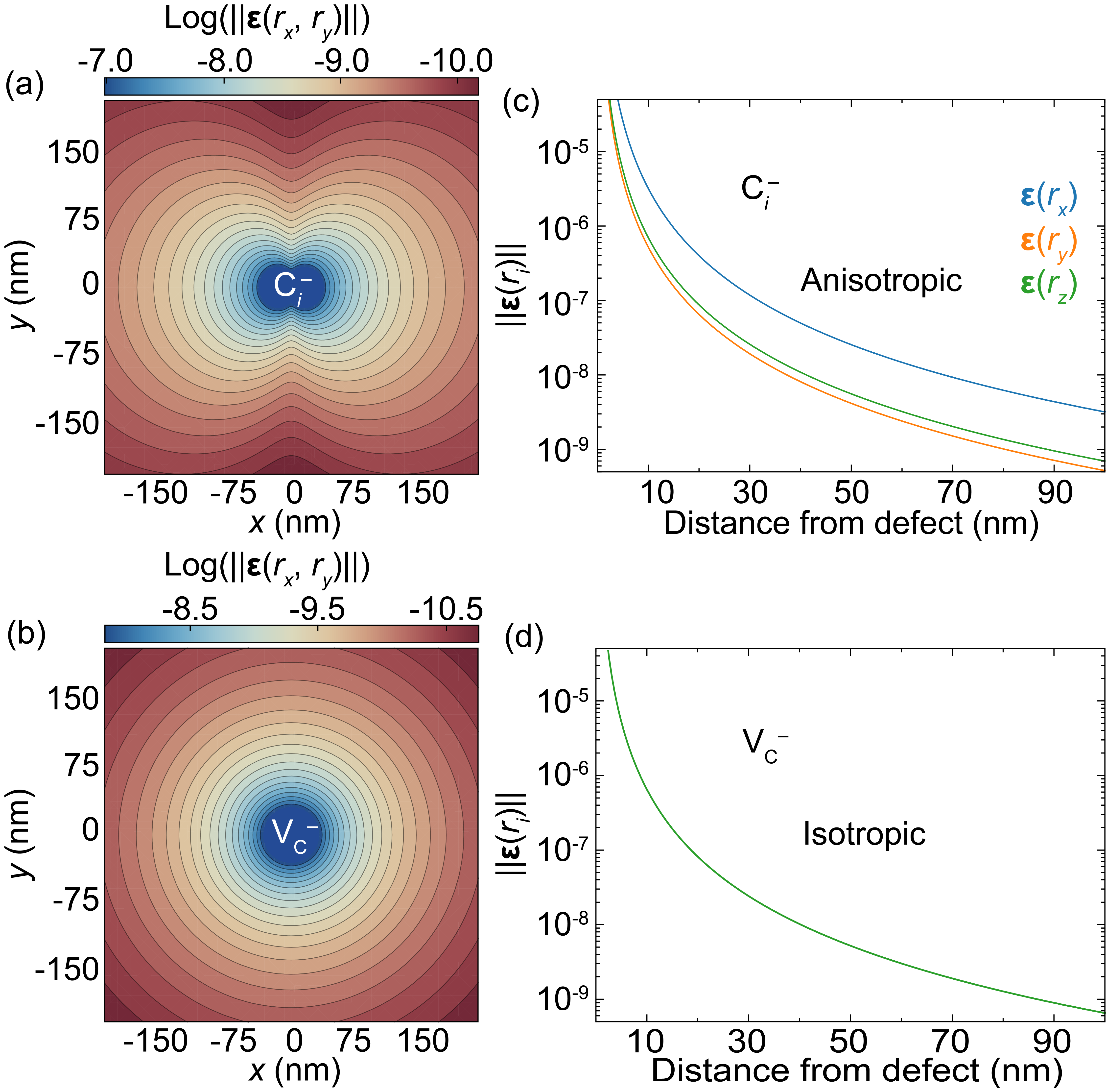}
\caption{Elastic dipole model representation of defect-induced strain fields in diamond. (a, b) 2D projection of the atomic strain field norms around the \ci{-} and \vc{-} defects, respectively. (c, d) 1D projection of the strain field norms along each Cartesian direction for \ci{-} and \vc{-} defects, respectively. In the case of \vc{-}, the three curves are almost identical, due to its isotropic nature.}
\label{fig:5}
\end{figure}

In analogy to the electric fields, was can also analyze the anisotropy of the strain resulting from the defects. Unlike the case of defect-induced electric fields, where the spherically symmetric contribution always dominates at any significant distance, defects can exhibit very anisotropic strain fields over longer distances. To demonstrate this, we take the comparison between negatively-charged \ci{} and the \vc{} as an illustrative example. As shown in Fig.~\ref{fig:5}{(a) and (b)}, the \ci{-} is highly anisotropic, resembling a dipole-like distortion field that points in the direction of the split-interstitial axis, while the case of \vc{-} is spherically symmetric and thus isotropic. This is all consistent with the results included in Fig.~\ref{fig:4}{(c)}; the spread of the scatter for \ci{-} is larger compared to that of \vc{-}, showcasing the higher symmetry/isotropy in the latter case. This can be further highlighted by comparing the individual Cartesian components of the strain fields in question [presented in Fig.~\ref{fig:5}{(c) and (d)}]. Clearly, the anisotropy of the \ci{-} defect propagates up to the far field. Thus, even at such distances, these two defects should generate drastically different strain fields, in contrast to their electric fields which originate from the same monopole charge and would thus be indistinguishable even in the near-field. As such, despite the fact that both electric and strain fields have symmetry-equivalent effects on the optical transition energies of the \nv{}~\cite{Rogers2009, Maze2011, Doherty2011, Doherty2013}, our results suggest that strain is a possible means to discriminate between intrinsic defects in diamond with the same charge state, assuming multiple \nv{} centers around the defect can be measured, see Sec.~\ref{sec:C2}.

\begin{figure}[h b]
\centering
\includegraphics[width=0.75\linewidth]{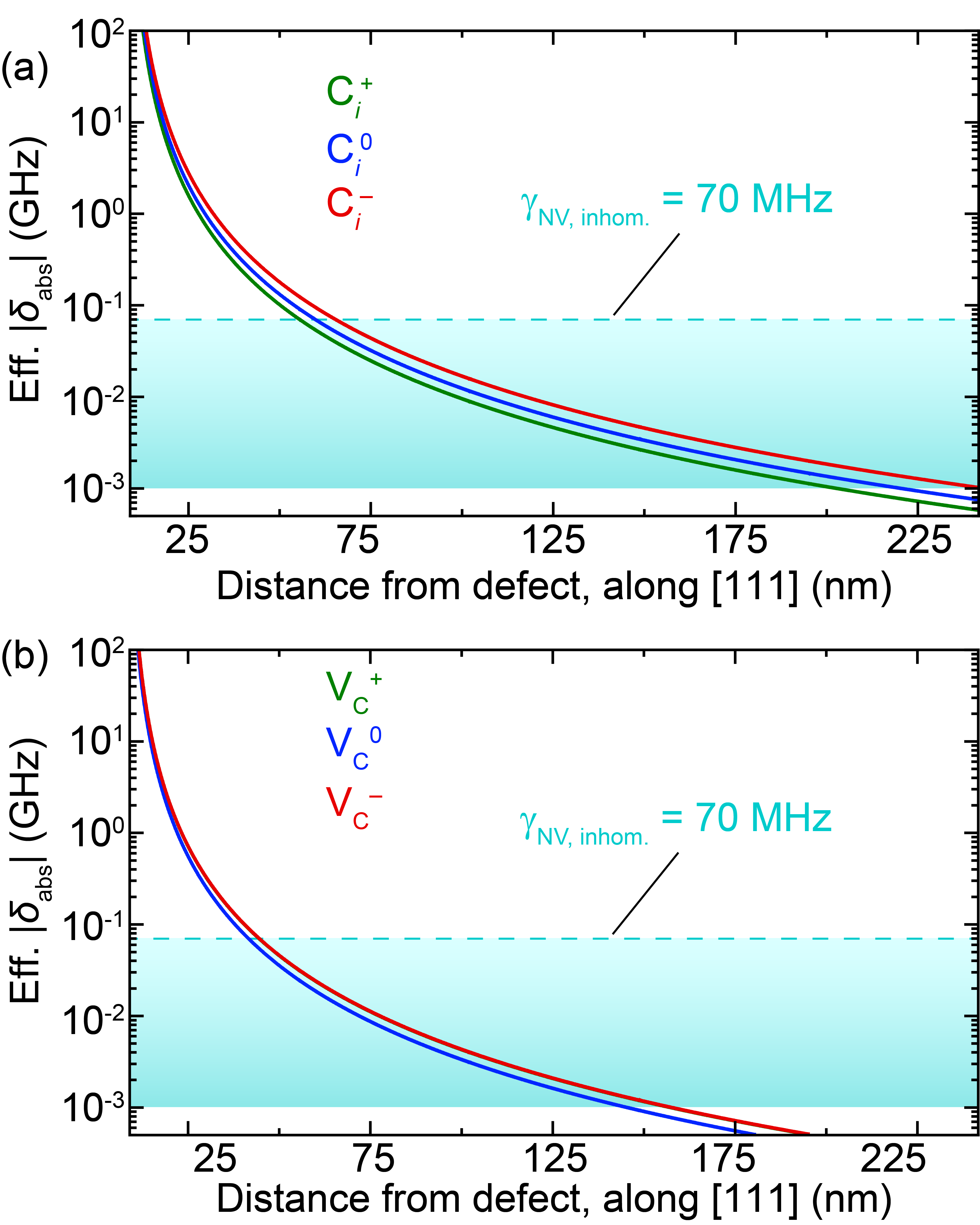}
\caption{Extrapolated spectral shifts on the \nv{} vertical transition based on the effective models in Fig.~\ref{fig:5} and the linear component of the longitudinal strain susceptibility $\chi_{_{A_1}}$. Horizontal dashed lines represent the detection limit based on the typical inhomogeneous (70 MHz) optical linewidth. Shaded regions highlight the resolution range based on the specifics of the experimental detection protocol.}
\label{fig:6}
\end{figure}

By projecting the strain tensor from a given native defect (i.e., \ci{} or \vc{}) along the \nv axis (here assumed to be [111]) direction, we obtain the component ``longitudinal'' to a hypothetically nearby \nv{} center at arbitrary distances away from it. This can be directly translated into strain-induced spectral shifts via the strain susceptibilities of the \nv optical excited states, see Sec.~\ref{sec:A}. Here, we focus on the longitudinal strain susceptibility $\chi_{_{A_{1}}}$ to obtain the spectral shifts on the vertical optical transition $\delta_{\text{abs}}$ induced by \ci{} or \vc{} defects. The results obtained from the $6\times6\times6$ defect supercells, extrapolated to experimentally realistic distances, are presented in Fig.~\ref{fig:6}. As in Fig.~\ref{fig:3}, we add the typical inhomogeneous linewidth of an \nv{} center in high purity diamond (70 MHz) as an indication of the minimum spectral shift experimentally resolvable. Analogous to the charge detection, the exact resolution will ultimately depend on the experimental protocol and averaging time~\cite{McCullian2022, Delord2024, Ji2024}, which we illustrate via the shaded regions in Fig.~\ref{fig:6}. Overall, the results from the elastic model suggest that individual \ci{} and \vc{} defects within a radius of $< 200$ nm and $< 150$ nm (respectively) should be within the sensing limit of a single \nv{} via strained-induced optical shifts. Furthermore, we see similar detection limits across the charged/neutral \ci{} and \vc{} defects, suggesting weak charge-state effects on the generated longitudinal strain fields. For simplicity, all these results derive from longitudinal components of the strain fields alone. With the relative anisotropy of some of these strain fields (e.g., \ci{-}), and the anisotropic nature of diamond, we expect measurements of strain-induced splitting of the $^{3}E$ states to improve these detection limits. Therefore, we conservatively estimate that these native defects would be detectable via their strain fields by an \nv{} up to 200 nm away. A more complete picture of both longitudinal and transverse local strains, along with the anisotropic evolution of the \nv excited states under such perturbations is considered in the following Section.

\begin{figure*}
\centering
\includegraphics[width=1.0\textwidth]{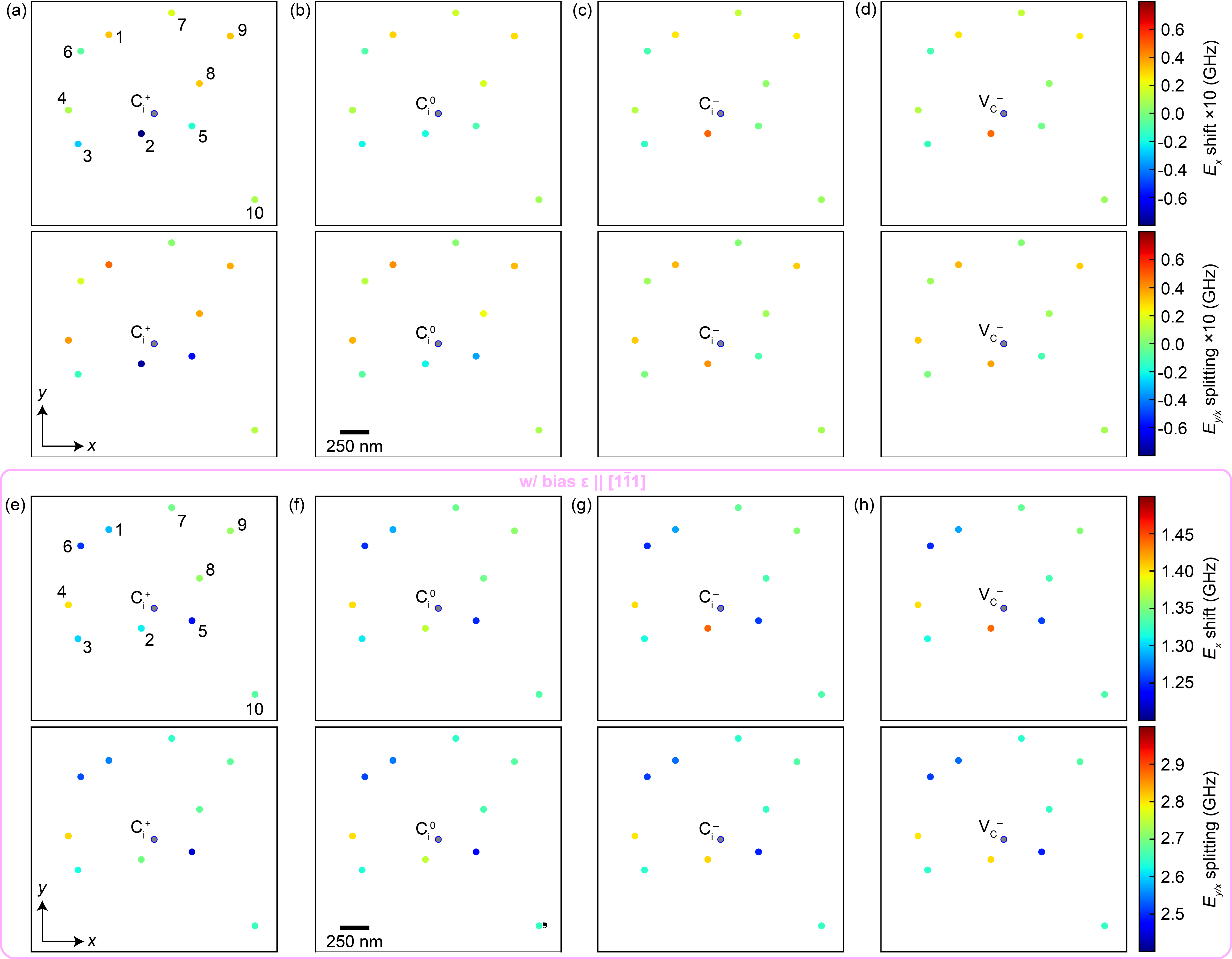}
\caption{Spectral response of an \nv cluster surrounding a single native defect in diamond. $^{3}E_{y}$ shift (top), $^{3}E_{x}$ shift (center), and $^{3}E_{y/x}$ splitting (bottom) across a random \nv distribution (kept fixed) surrounding a single (a) \ci{+}, (b) \ci{0}, (c) \ci{-}, and (d) \vc{-} defect center. All results are based on the individual $^{3}E_{y/x}$ strain susceptibilities and permanent dipoles of the \nv center.}
\label{fig:7}
\end{figure*}

\subsection{Measurements on multiple \nv{} centers for characterizing native defects \label{sec:C2}}
In real experiments, the effects described independently in Secs.~\ref{sec:A2} and~\ref{sec:B2} interact to yield complex perturbation fields that propagate across diamond. The results in the previous sections can be used to determine quantitatively what the effect of a nearby native defect will be on the \nv{} given the defect type, charge state, and position/orientation relative to the \nv{}. As mentioned above, we could also use this information to directly probe native defects (or impurities) present in the crystal via the \nv{} center. The idea here would be to measure the perturbations of the properties of \emph{multiple} \nv{} centers in the vicinity of a defect. In fact, it has been demonstrated that small ensembles of \nv{} centers can actually be monitored \emph{simultaneously}~\cite{Chen2019, Ji2024, Delord2024, Guo2024}. We note that because of the stray strain and electric fields that are always present in diamond samples~\cite{Tamarat2006, Bassett2011, Acosta2012, Lee2016, McCullian2022}, the absolute magnitudes of the deviations from ``ideal'' \nv{} spectra may be of limited use; it may be more useful to measure \emph{changes} with the creation of defects, e.g., via controlled irradiation or annealing. In order to explore the efficacy of these ideas, we combine the results outlined in previous sections with a hypothetical random distribution of \nv{} centers around a single native defect in diamond, see Sec.~\ref{sec:multi_NV} for details.

The results of an example simulation are summarized in Fig.~\ref{fig:7}. The top panels present the $^{3}E_{x}$ shifts and the $^{3}E_{y} \leftrightarrow$ $^{3}E_{x}$ splitting for all three charge states of the \ci{} [Figs.~\ref{fig:7}{(a--c)}] and for \vc{-} [Fig.~\ref{fig:7}{(d)}]. Within these calculations, we consistently find that electric-field-induced effects dominate over the strain ones, which on average are 1--2 orders of magnitude smaller. This is expected from the results in Secs.~\ref{sec:A2} and~\ref{sec:B2}, since the strain fields decay much faster than electric fields. Comparing across Figs.~\ref{fig:7}{(a--c)}, we see that in the $< 500$ nm vicinity of a \ci{}, a modest cluster of 3 \nv centers could suffice to spectroscopically differentiate between its different charge states, well within the working distances of recent triangulation experiments~\cite{Li2024, Delord2024}. On the other hand, Figs.~\ref{fig:7}{(c)} and~\ref{fig:7}{(d)}, suggest that optical discrimination between \ci{-} and \vc{-} may be more challenging; the $\propto 1/r^{3}$ decay of strain renders the electric-field effects dominant over such distances, making defects with the same charge monopole indistinguishable. However, within a 10--30 nm radius around the defects, their drastically different strain fields (Fig.~\ref{fig:5}), should induce spectral effects comparable to those from their charges, making them distinguishable. This tighter constraint on the NV-cluster sensor may prove challenging experimentally, however defect-charge neutralization could be used to resolve the strain effects up to a longer range~\cite{monge_beyond_2025}.

Under the influence of a background macro/mesoscopic strain field, which is typically present in real experiments, the situation is more stringent [Figs.~\ref{fig:7}{(e--h)}]. Here, we have arbitrarily considered a diagonal strain field along the [1$\bar{1}$1] direction, with $|\bm{\varepsilon}| = 1\times10^{-7}$ chosen based on the induced splittings of $\sim 3$ GHz. In this case, only the \nv closest to the defect center (``NV 2'') seems to be reliably affected by a nearby \ci{} with different charge states. The same conclusions derived from comparing Figs.~\ref{fig:7}{(c)} and~\ref{fig:7}{(d)} can be drawn from Figs.~\ref{fig:7}{(g)} and~\ref{fig:7}{(h)}. Overall, though the specific experimental conditions (background strain, spectral inhomogeneities, noise level, etc.) will play a role in determining how many NVs surrounding a single native defect can be used collectively to reliably sense its perturbations, our results suggest that a small clusters of 2--5 \nv centers could be enough to collectively sense and discriminate nearby native defects, in line with recent experimental work~\cite{Delord2024, Ji2024}.

\begin{figure*}
\centering
\includegraphics[width=1.0\textwidth]{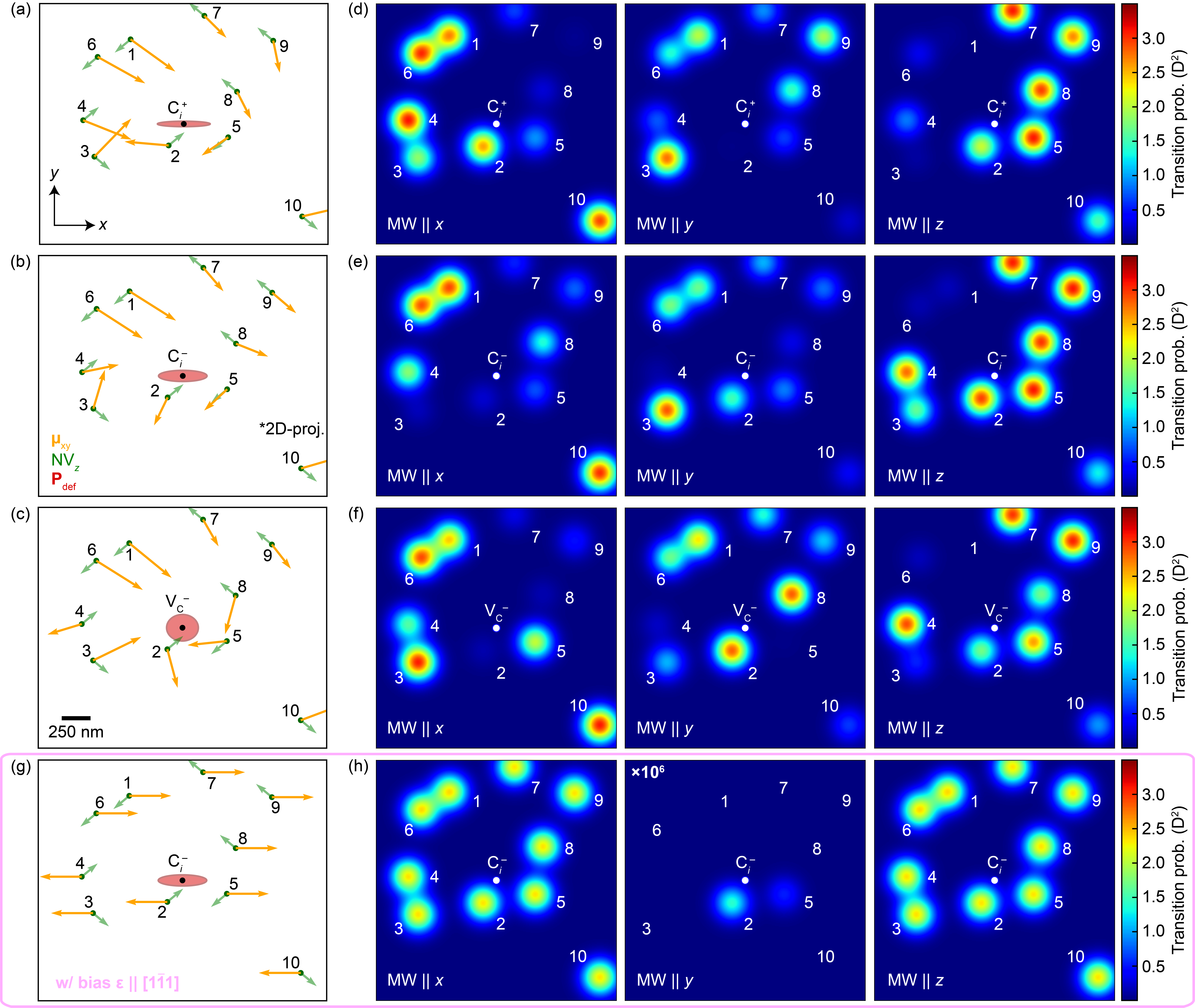}
\caption{Local effects on the optical excited states of a random \nv{} distribution around hypothetical native defects. (a) 2D projection of the random \nv orientations, along with the alignment of their $^{3}E_{x} \leftrightarrow$ $^{3}E_{y}$ transition dipole moment ($\bm{\mu}_{xy}$). An effective (projected) elastic dipole tensor (here denoted as ``$\textbf{P}_{\text{def}}$'') is also included for reference. (b) Optical maps of the \nv distribution under broadband MW excitation for different MW polarizations, based on the $^{3}E_{x} \leftrightarrow$ $^{3}E_{y}$ selection rules under strain. The results in (g, h) include an arbitrary bias (background) strain field along the [1$\bar{1}$1] direction.}
\label{fig:8}
\end{figure*}

As discussed in Sec.~\ref{sec:multi_NV}, a complimentary measurement to the shifts/splittings outlined in Fig.~\ref{fig:7}, is of the transition strength between $^{3}E_{y/x}$ excited states of the \nv{}, and its alignment with the local perturbations. The obtained $\bm{\mu}_{xy}$ for each \nv (projected in the $xy$ plane) are presented in Figs.~\ref{fig:8}{(a--c)}. To demonstrate the working principle, we imagine a MW drive tuned to the $^{3}E_{x} \leftrightarrow$ $^{3}E_{y}$ transition, but with a polarization that can be changed to point in the three Cartesian directions. In this case, the intensity of the transition will depend on the projection of $\bm{\mu}_{xy}$ along the polarization direction. Using this idea, we plot in Fig.~\ref{fig:8}{(d--f)} modulated 2D excitation maps for the \nv{}/native defect structures shown in in Fig.~\ref{fig:8}{(a--c)}. The normalized intensity is scaled by the magnitude of the projected dipole moment on the given polarization direction, and broadened by Gaussian distributions of 100-nm width to mimic the spatial resolution of an optical measurement of single \nv{} centers. For this example situation, we see that the $\hat{x}$ and $\hat{y}$ MW polarizations provide a clear distinction between the three defects, demonstrating that it could be a way to differentiate between different defects of like charge states. 

There are a few caveats to these proposed experiments. Firstly, they are quite sensitive to background strain, as we show in Fig.~\ref{fig:8}(g) and (h), which have the same 3-GHz $[1\overline11$] strain as in Fig.~\ref{fig:7}(e--h). In this case, all of the dipoles are aligned with the background strain, and are not sensitive to the type/charge state of the defect. Also, these calculations assume resonant MW excitation at each \nv center, which will likely pose experimental challenges. Additionally, rotating the MW polarization in real experiments may be inconvenient. However, as described in Sec.~\ref{sec:B}, the selection rules derived for the $^{3}E_{x} \leftrightarrow$ $^{3}E_{y}$ MW transition are analogous to those for the individual $^{3}A_{2} \leftrightarrow$ $^{3}E_{x/y}$ optical transitions~\cite{Fu2009, Maze2011, Doherty2011, Bassett2011, Happacher2022}, i.e., the proposed measurements should also work via polarization protocols on the resonant optical excitation beam. In addition, we have only considered a single native defect in the vicinity of the \nv centers. We expect the more realistic situation of multiple native defects to be simply a linear superposition of strain and electric fields. Thus, we propose such MW/optical polarization experiments as a valuable spectroscopic tool complementary to other (e.g., photoluminescence excitation) techniques for detecting/discerning native defects in diamond, or background strains in general.

\section{Summary and Conclusions \label{sec:conclusions}} 
We have used first-principles calculations based on density-functional theory to study the effects of native defects on nearby \nv{} centers in diamond. The short range strain and electric fields of the defects are calculated with DFT, and then extrapolated with continuum models to the $\sim 1$ $\mu$m length scales relevant for experiments. We then combined these results with quantum embedding calculations of \nv{} strain and electric field susceptibilities~\cite{Lopez-Morales2024} and performed simulations where experimentally relevant concentration of \nv{} centers were in the vicinity of a native defect. We find that the electric fields from charged native defects have a measurable effect on the \nv{} optical spectra over a micron away, while for neutral defects it could be up to 100s of nm. As expected the main contribution in the charged case is the spherically symmetric monopole term, which dominated even at very short distances of just a few nanometers. The strain fields of the native defects have a much smaller extent, resulting in detectable changes to the \nv{} up to $\sim200$ nm away. In addition, depending on the structure of the defect, the strain field may be significantly asymmetric.

Using these results, we demonstrate the utility of multi-\nv{} measurements on characterizing nearby native defects in diamond. We see that at \nv{} concentrations of 2.5 NV/$\mu$m$^2$, there are often a handful of \nv{} centers in the immediate vicinity of the native defect. The position of the native defect can be determined by the decay in strain and electric-field related perturbations to the \nv{} spectra. Also, the anisotropy in the strain field is an important way of differentiating between electric-field and strain effects, allowing to identify different defects, and even different charge states of the same defect. Additional utility is provided by the directional transition dipoles between the strain-split excited states, which can be probed by, e.g., changing the polarization of a MW drive.

These calculations not only provide a quantitative picture of how defects affect the optical properties of the \nv{}, but also validate the idea that \nv{} centers can be used to detect and characterize defects in diamond through multi-\nv{} measurements such as those simulated in this work. This could have applications in materials characterization, radiation detection, or even, as mentioned in the introduction, detection of high-energy particles. We also believe it to be easily generalizable to other insulating materials with well-described color centers, such as, silicon, silicon carbide, or boron nitride. Therefore, the door is open for color-center quantum metrology to be applied not only for exploring external stimuli, but also internal perturbations and the structure of the host material itself.

\acknowledgements
CED acknowledges support from the National Science Foundation under Grant No.~DMR-2237674. GILM acknowledges funding from Grant NSF-2208863. TD and CAM acknowledge support from the National Science Foundation through grants NSF-2203904 and NSF-1914945 and from the U.S. Department of Energy, Office of Science, National Quantum Information Science Research Centers, Co-design Center for Quantum Advantage (C2QA) under contract number DESC-0012704. JMZ would like to acknowledge BNL-SBU Seed Grant funding. The Flatiron Institute is a division of the Simons Foundation.

\bibliography{NV_bib}
\end{document}